\documentclass[reprint, superscriptaddress, amsmath,amssymb, aps, prb, floatfix, longbibliography]{revtex4-2}

\usepackage{amsfonts}
\usepackage{braket}
\usepackage{dsfont}
\usepackage{graphicx}
\usepackage{xcolor}
\usepackage{listings}
\usepackage{comment}
\usepackage{hyperref}
\usepackage{booktabs} 

\usepackage{footnote}
\makesavenoteenv{figure}

\usepackage[caption=false]{subfig}

\begin{document}

\preprint{APS/123-QED}

\title{Assessing Majorana states and qubits through quantum capacitance}

\author{Rodrigo A. Dourado}
\email{dourado.rodrigo.a@gmail.com}
\affiliation{ \textit{Instituto de F\'{ı}sica de S\~{a}o Carlos, Universidade de S\~{a}o Paulo, 13560-970 S\~{a}o Carlos, S\~{a}o Paulo, Brazil}	}

\author{Ramón Aguado}
\affiliation{Quantum Advanced Research Center (QuARC), Consejo Superior de Investigaciones Científicas (CSIC), 28049 Madrid, Spain}
\affiliation{Instituto de Ciencia de Materiales de Madrid (ICMM), Consejo Superior de Investigaciones Cient\'{i}ficas (CSIC), 28049 Madrid, Spain}

\author{Jeroen Danon}
\affiliation{Department of Physics, Norwegian University of Science and Technology, Trondheim NO-7491, Norway}

\author{Martin Leijnse}
\affiliation{Division of Solid State Physics and NanoLund, Lund University, S-22100 Lund, Sweden}

\author{Rub\'{e}n Seoane Souto}
\affiliation{Quantum Advanced Research Center (QuARC), Consejo Superior de Investigaciones Científicas (CSIC), 28049 Madrid, Spain}
\affiliation{Instituto de Ciencia de Materiales de Madrid (ICMM), Consejo Superior de Investigaciones Cient\'{i}ficas (CSIC), 28049 Madrid, Spain}

\begin{abstract}
Quantum capacitance (QC) has recently emerged as a promising tool for parity readout in topological qubits based on Majorana bound states (MBSs). Here, we show that this capability can be extended further: by employing an auxiliary quantum dot (QD) as a sensor, we demonstrate that QC measurements simultaneously resolve two fundamental figures of merit of the device, the ground-state energy splitting and the MBS overlap, thus providing direct access to the underlying internal degrees of freedom. Using a low-energy effective model, we provide analytic expressions for these two figures of merit that can be determined from the relative position and magnitude of the QC maxima in the even and odd parity sectors as functions of the auxiliary-QD energy. We further validate these results with a microscopic model of QD-based Kitaev chains and qubits, demonstrating their applicability in a wide range of MBS-based devices. Our results establish QC as a probe of MBS quality and a tool for topological-device optimization that preserves fermion parity. 
\end{abstract}
	
\maketitle

\textit{Introduction.} Majorana bound states (MBSs) are promising for realizing quantum devices intrinsically protected against local sources of decoherence through a non-local encoding of quantum information~\cite{Kitaev2001,kitaev2003fault}. In particular, MBS-based qubits exploit fermion parity encoded non-locally as a protected degree of freedom~\cite{nayak2008non}. Over the past decade, significant progress has been achieved in engineering hybrid semiconductor–superconductor systems designed to host and manipulate such states in nanowires~\cite{lutchyn2010majorana, oreg2010helical, mourik2012signatures, beenakker2020search, flensberg2021engineered, aguado2017majorana, alicea2012new, prada2020andreev, pitavidal2025novelqubitshybridsemiconductorsuperconductor} and short arrays of quantum dots (QDs)~\cite{Dvir2023, bordin2023crossed, bordin2024signatures,Bordin_arXiv2025, haaf2024edgebulkstatesthreesite, van2024cross, zhang2025gatereflectometryminimalkitaev, wang2023triplet, ten2024two, Souto_chapter}.

Due to the exotic properties of MBSs, including their non-Abelian statistics and topological protection, several protocols have been developed to distinguish MBSs from trivial Andreev bound states \cite{prada2020andreev}, including local tunneling spectroscopy, three-terminal nonlocal conductance, and topological-gap protocols~\cite{gul2018ballistic, pan2021quantized, Pikulin_arXiv2021, Aghaee_PRB2023}. A particularly direct approach consists of coupling the system to an auxiliary QD, which can convert MBS overlap into a splitting of the even--odd ground-state degeneracy
~\cite{Deng_Science2016,Prada_PRB2017, Bordin_arXiv2025, Clarke_PRB2017,PhysRevB.98.085125,Seoane2023}. However, these characterization schemes rely on electron exchange with an external reservoir, thereby inducing transitions between different fermion-parity states, which rules them out as valid experimental probes of MBS coherent properties. Furthermore, the microscopic details of the tunneling region complicates the physical interpretation~\cite{Lee2014,PhysRevB.98.235406,Vuik_SciPost19, Avila_ComPhys2019}.

Readout of the quantum state encoded in MBSs rely on converting the non-local fermion parity into a measurable signal by accessing both MBSs simultaneously. Charge-based schemes achieve this either through parity-to-charge conversion using tunnel-coupled QDs~\cite{Gharavi_PRB2016,Szechenyi_PRB2020,Munk_PRR2020,Steiner_PRR2020,Schulenborg_PRB2021}, or through dispersive coupling to an RF resonator, where the measured signal is a parity-dependent susceptibility or quantum capacitance (QC)~\cite{Smith_PRXQ2020,peri2024unified,liu2026quantumcapacitanceparityswitching}. Recent experiments have demonstrated that QC can act as a global probe of MBS parity, enabling time-resolved single-shot readout while local charge sensing remains insensitive to the state encoded in MBSs due to their non-local character~\cite{microsoft2025interferometric, vanloo2025singleshotparityreadoutminimal, zhang2025gatereflectometryminimalkitaev}. These developments make capacitive and reflectometry-based charge sensing a particularly suitable platform for MBS readout, provided quasiparticle poisoning and measurement-induced back-action remain sufficiently suppressed.

In this work, we demonstrate a different use of QC, showing that when measured at an auxiliary QD, it provides direct information about two key metrics of MBS devices: the local MBS overlap and the ground-state energy splitting. Using a minimal model, we demonstrate that the position and relative magnitude of the QC maxima in the even and odd parity sectors, as functions of the auxiliary-QD energy, directly correlate with the even--odd energy splitting and the local MBS overlap at the point where the auxiliary QD couples to the system. We validate these results with a microscopic model of QD-based Kitaev chains~\cite{Sau2012, fulga2013adaptive, tsintzis2022creating, Tsintzis2024, dourado2025twositekitaevsweetspots, Dourado_Kitaev3,Alvarado_PRB2024,Alvarado2026}, recently realized experimentally~\cite{Dvir2023, ten2024two, Haaf2023, Bordin2024, haaf2024edgebulkstatesthreesite, vanloo2025singleshotparityreadoutminimal, zhang2025gatereflectometryminimalkitaev}, and show that our conclusions remain applicable to longer systems, including MBS-based qubits~\cite{Pino_PRB2024, pan2025rabi}.

\begin{figure}
\centering
\includegraphics[width=0.9\linewidth]{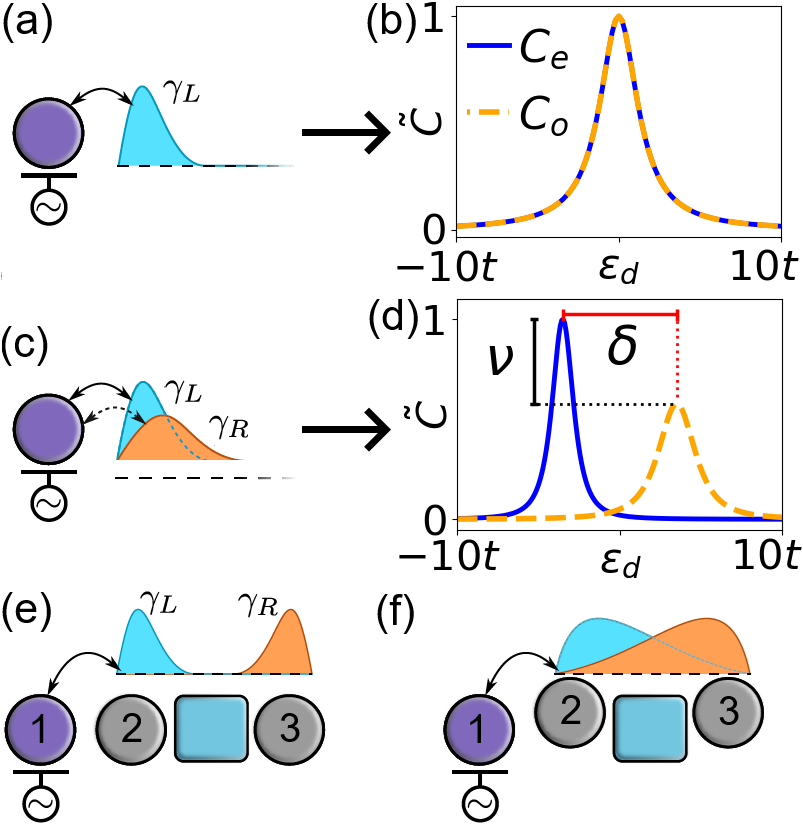}
\caption{Identifying MBSs via QC. (a) A zero-energy MBS coupled to a QD, itself coupled to a resonator, produces (b) identical QC responses in the even and odd parity sectors as a function of $\varepsilon_d$. (c) Overlapping MBSs ($u^2 = 0.75$, $v^2 = 0.25$) at finite energy coupled to a QD lead to (d) split QC peaks, represented by $\delta$, with unequal amplitudes, quantified by $\nu$, in the even and odd parity sectors. (e)-(f) Schematic representation of QD-based Kitaev chains hosting, respectively, localized and delocalized MBSs.}
\label{Fig1}
\end{figure}

\textit{Measurement of energy splittings and MBS overlaps using QC.} To understand the working principle of our proposed method, we consider a QD coupled to an isolated or two overlapping MBSs, Figs.~\ref{Fig1}(a,c). The QD, in turn, couples to a resonator, from which the QC can be measured. The MBSs, described by the operators $ \gamma_{L,R}$, form together a fermionic mode $f = \frac{1}{2}(\gamma_L + i\gamma_R)$, with $\gamma_i^\dagger=\gamma_i$ and $\{\gamma_i,\gamma_j\}=2\delta_{ij}$. The effective low-energy Hamiltonian that describes the couplings in Figs.~\ref{Fig1}(a,c) can be written as~\cite{Clarke_PRB2017}
\begin{equation}
\label{Eq:simple_H}
H = \varepsilon_d\, d^\dagger d + \varepsilon_M\, f^\dagger f
+ \left[\, d^\dagger \left(  t_L \gamma_L +i t_R \gamma_R \right) +\mathrm{H.c.} \right],
\end{equation}
where $\varepsilon_d$ is the QD energy level, $\varepsilon_M$ is the MBS hybridization energy, and $t_{L,R}$ are the coupling to the two MBSs. It is convenient to rewrite the hybridization Hamiltonian, last term in Eq.~\eqref{Eq:simple_H}, as $\left[t\, d^\dagger \left( u f + v f^\dagger \right) + \mathrm{H.c.} \right]$, with $t=\sqrt{2(t_L^2+t_R^2)}$ and the coherence factors given by $u=(t_L+t_R)/t$ and $v=(t_L-t_R)/t$.
In particular, the situation in Fig.~\ref{Fig1}(a), with well-localized MBSs, where QD couples to a single MBS, is characterized by $u = v$. Deviations from this case, ($u \neq v$), describe overlapping MBSs, Fig.~\ref{Fig1}(c). 

Since the above Hamiltonian conserves fermion parity, the system can be divided into separate sectors, enabling the calculation of the curvature of the ground-state energies with respect to $\varepsilon_d$, which is related to the QC. Although the total fermionic parity remains conserved, the parity of the individual components, namely, the QD and the fermionic mode $f$, can change simultaneously due to electron exchange processes. These events manifest as peaks in the QCs and provide information about the energy splitting and overlap of the MBSs. The QCs can be obtained by diagonalizing the Hamiltonian, and are given by
\begin{equation} \label{Ce}
C_{\mathrm{e}}
=
-\frac{\partial^2 E_{0}^{\mathrm{e}}}{\partial \varepsilon_d^2}
=
\frac{2|t v|^2}
{\left[(\varepsilon_d+\varepsilon_M)^2 + 4|t v|^2\right]^{3/2}},
\end{equation}
\begin{equation} \label{Co}
C_{\mathrm{o}}
=
-\frac{\partial^2 E_{0}^{\mathrm{o}}}{\partial \varepsilon_d^2}
=
\frac{2|t u|^2}
{\left[(\varepsilon_d-\varepsilon_M)^2 + 4|t u|^2\right]^{3/2}},
\end{equation}
where $E_0^{(e,o)}$, represents the ground state energy of the even and odd parity sectors, respectively. We now consider a few limiting cases. For well-localized MBSs with no hybridization, i.e., $\varepsilon_M = 0$ and $u = v$, see Fig.~\ref{Fig1}(a), $C_e = C_o$ for any $\varepsilon_d$, as shown in Fig.~\ref{Fig1}(b). Deviations from this ideal scenario cause distinctive signatures in the QCs. For instance, by setting $\varepsilon_M \neq 0$, the QC in the odd and even states are shifted in $\varepsilon_d$ by $\pm \varepsilon_M$, respectively. The split between the peaks in the QCs, which we define as $\delta$, can be used to measure the energy splitting through the relation $\varepsilon_M = \delta/2$. Aside from the energy splitting, noting that the maximum values of the QCs in the even and odd sectors are determined by $v$ and $u$, respectively, we can also quantify the MBS overlap through the ratio of the QCs' peaks. We define the quantity $\nu = |C_e^{\rm max} - C_o^{\rm max}|/C_{\rm max}$, where $C_{\rm max}$ is the maximum value of the QCs. In terms of the coherence factors,

\begin{equation} \label{nuDefinition}
\nu = 1 - \frac{\min(|u|,|v|)}{\max(|u|,|v|)},
\end{equation}
which allows us to quantify the local MBS overlap through the Majorana polarization (MP)~\cite{sticlet2012spin, tsintzis2022creating, sedlmayr2015visualizing, sedlmayr2016majorana, dourado2025twositekitaevsweetspots, douradoCMA2026}, defined as
\begin{equation} \label{MPdef}
M_i = \frac{|w_i|^2 - |z_i|^2}{|w_i|^2 + |z_i|^2} = \frac{2 |u_i| |v_i|}{|u_i|^2 + |v_i|^2},
\end{equation}
where the middle expression is written in the many-body representation, with $w_i = \bra{E} (c_i + c_i^\dagger) \ket{O}$ and $z_i = \bra{E} (c_i - c_i^\dagger) \ket{O}$, where $\ket{E}$ and $\ket{O}$ denote the ground states in the even and odd parity sectors, respectively. Well-localized MBSs, with zero overlap, lead to $|M_i|=1$ at the ends of the system, while a finite overlap between MBSs reduces MP, $|M_i|<1$. The rightmost expression in Eq.~\eqref{MPdef} is written in the single-particle representation, where $u_i$ and $v_i$ are the local coherence factors of the Bogoliubov-de Gennes wave functions. From Eqs.~(\ref{nuDefinition}) and (\ref{MPdef}), one can show that
\begin{equation} \label{MPestimate}
|M_i| = \frac{2(1 - \nu)}{1 + (1-\nu)^2},
\end{equation}
where $\nu$ is measured in an auxiliary QD attached to site $i$.

In summary, overlapping MBSs lead to different peak QC values in the two parity sectors, quantified by the parameter $\nu$, from which the MP can be estimated that directly relates to the local overlap between MBSs. Energy splittings can be extracted from the QC peak splitting, $\delta$. In general, both effects may occur simultaneously, resulting in QC profiles with unequal and split peaks in the even and odd parity sectors, as illustrated in Fig.~\ref{Fig1}(d). Therefore, tuning the system toward well-localized zero-energy MBSs requires minimizing both $\delta$ and $\nu$. In experiments, the QC in the two parity sectors can be measured regardless of whether the quasiparticle poisoning time is shorter or longer than the QD sweep time: if the poisoning time is shorter, both parity sectors are sampled within a single sweep, whereas if it is longer, a separate sweep for each parity sector is sufficient.

Our QC method enables the quantification of MBS overlap and energy splitting, two key figures of merit that are difficult to extract using alternative transport-based approaches, such as transport through an auxiliary QD~\cite{Clarke_PRB2017, Prada_PRB2017, Seoane2023}. Furthermore, the method preserves parity, making it compatible with experiments designed to probe MBS coherence.
We further note that the above derivation is system-agnostic and should apply to any platform capable of hosting MBSs, including interacting systems~\cite{samuelson2025quantifyingrobustnesslocalitymajorana}.

\textit{QD-based Kitaev chains.} So far, we have shown that the QC can be used to quantify relevant figures of merit, such as the energy splitting and the MBS overlap, using a low-energy effective Hamiltonian to describe the MBSs. We now apply the above-discussed concepts to a particular model. We consider QD-based realizations of Kitaev chains as a concrete platform capable of hosting MBSs. In these systems, semiconducting QDs are coupled through superconducting segments, shown in gray and blue, respectively, in Fig.~\ref{Fig1}(e). The interplay between superconductivity, spin-orbit coupling, and an applied magnetic field can lead to the emergence of MBSs. We initially consider the limit of sufficiently large spin splittings and weak QD-superconductor coupling such that the system depicted in Fig.~\ref{Fig1}(e) can be mapped onto an effective two-site Kitaev chain~\cite{liu2022tunable, Kitaev2001}. We couple the Kitaev chain to an auxiliary QD on the left side, see Fig.~\ref{Fig1}(e). The Hamiltonian of the system reads

\begin{equation}
H_K = \sum_{i=1}^3 \varepsilon_i c_i^\dagger c_i + \left[\sum_{i=1}^2 t_{i, i+1} c_{i}^\dagger c_{i+1} + \Delta_{2, 3} c_{2}^\dagger c_3^\dagger + \text{H.c.} \right],
\end{equation}
where $c_i$ ($c_i^\dagger$) annihilates (creates) a spinless electron at site $i$, $t_{(i, i+1)}$ denotes the hopping amplitude between neighboring sites, and $\Delta_{2,3}$ is the effective p-wave pairing amplitude. The hopping $t_{2,3}$ and pairing $\Delta_{2,3}$ arise from elastic cotunneling and crossed Andreev reflection between the normal QDs, mediated by the superconducting segment~\cite{liu2022tunable}. 

The above-described minimal Kitaev chain supports fully localized MBSs, one at each end of the Kitaev chain, at the so-called \textit{sweet spots}~\cite{Leijnse2012, tsintzis2022creating}. To assess the quality of the MBSs, we evaluate the MP, Eq.~\ref{MPdef}, and the energy splitting, $|\delta E_0| = |E_0^e - E_0^o|$, where $E_0^{e}$ and $E_0^{o}$ are the ground-state energies in the even and odd fermion parity sectors, respectively.
The sweet spots, characterized by degenerate even and odd ground states ($\delta E_0 = 0$) and fully localized MBSs at the ends of the Kitaev chain ($|M_2| = |M_3| = 1$), occur at $\varepsilon_2 = \varepsilon_3 = 0$ and $t_{2,3} = \pm \Delta_{2,3}$~\cite{Leijnse2012} (all energies are expressed in units of the $\Delta_{2,3}$).

\begin{figure}
\centering
\includegraphics[width=\linewidth]{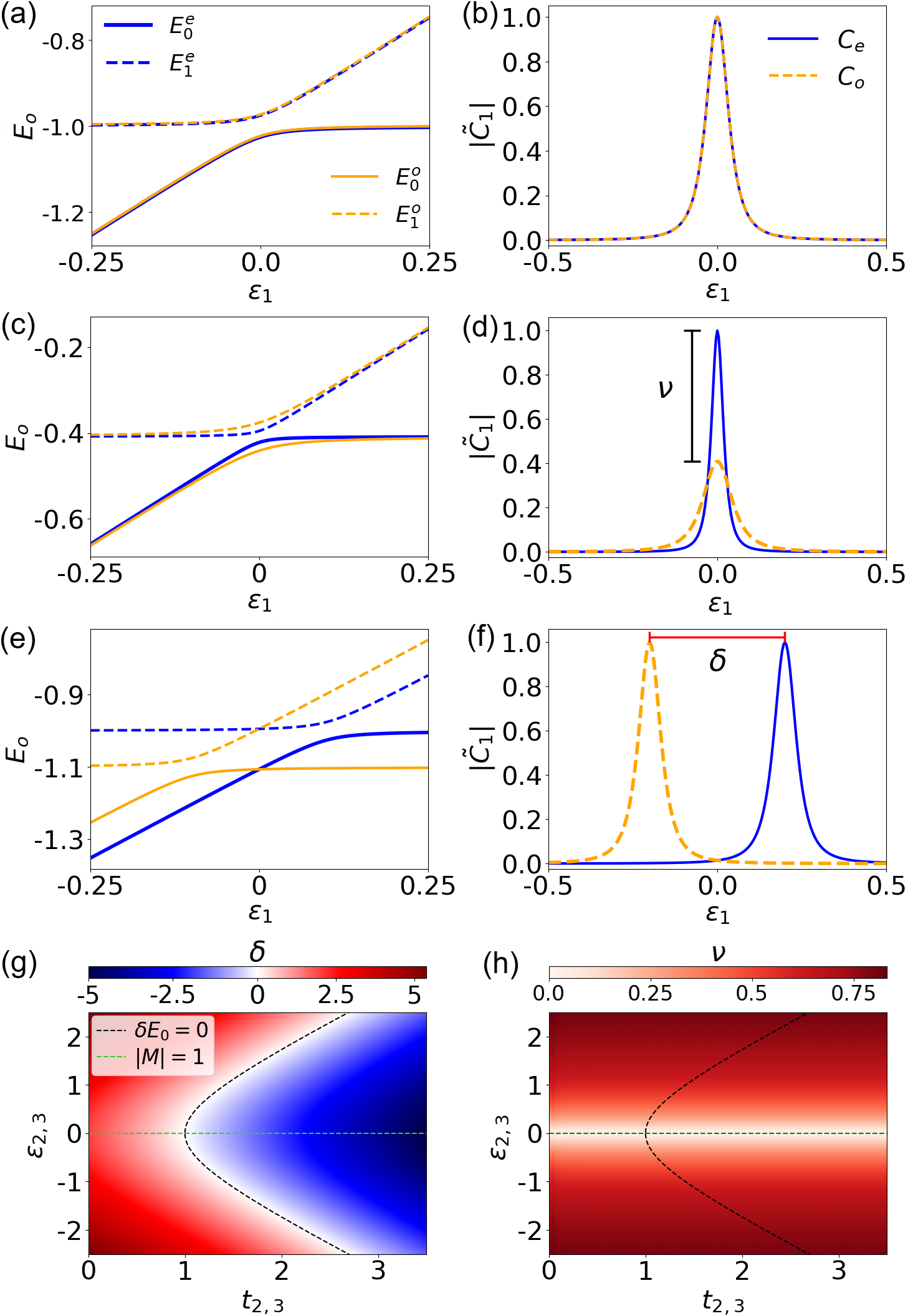}
\caption{Detecting MBS sweet spots in QD-based Kitaev chains.
(a),(b) MBS sweet spot with $|\delta E_0| = 0$ and $|M| = 1$ ($\varepsilon_{2, 3} = 0$, $t_{2, 3} = 1$), exhibiting degeneracy in both the energy spectrum (blue and green denote even and odd parity sectors, respectively) and the QC as the left QD level is varied. (c),(d) Poorly localized MBSs, $|M| = 0.7$ ($\varepsilon_{2, 3} = 1.02$, $t_{2, 3} = 1.43$), where different curvatures of the even and odd sectors lead to unequal QC peak heights. (e),(f) Finite energy splitting, $|\delta E_0| = 0.2$ ($\varepsilon_{2, 3} = 0$, $t_{2, 3} = 1.2$), resulting in split QC peaks. (g),(h) Peak splitting $\delta$ and height difference $\nu$ in the QC between even and odd parity sectors, as functions of $\varepsilon_{2, 3}$ and $t_{2,3}$. In all panels, $t_{1,2} = 0.05$.}
\label{Fig2}
\end{figure}

\textit{Identifying MBS sweet spots with QC in Kitaev chains.} We now apply the concepts discussed above to identify MBS sweet spots in Kitaev chains. We start our analysis by considering symmetric Kitaev chains, where $\varepsilon_2 = \varepsilon_3 = \varepsilon_{2, 3}$, which implies $|M_2| = |M_3| = |M|$, although we relax this constraint later. Our goal is to estimate the energy splitting and MP through QC measurements.

At the MBS sweet spot, $t_{2,3} = 1$ and $\varepsilon_{2,3} = 0$, the lowest energy levels in the even and odd parity sectors are degenerate for any $\varepsilon_1$, as shown in Fig.~\ref{Fig2}(a), resulting in identical QCs with peaks around $\varepsilon_1 = 0$, see Fig.~\ref{Fig2}(b). The QC peaks originate from anticrossings in the energy spectrum, signaling electron transfer between the QD and the Kitaev chain. This result corroborates our previous discussion of the effective model, where an isolated MBS coupled to a QD leads to equal QCs in the even and odd sectors, see Fig.~\ref{Fig1}(b).

We now move away from the sweet spot and consider poorly localized MBSs, $|M| = 0.7$, at zero energy, for which $u \neq v$. As shown in Fig.~\ref{Fig2}(c), the overlap between the MBSs breaks the symmetry between the even and odd parity sectors, leading to different curvatures of the energy levels and, consequently, to unequal QC peak heights in Fig.~\ref{Fig2}(d). In contrast to the ideal case [Fig.~\ref{Fig2}(b)], where electrons can be added or removed at no energy cost, the overlap lifts the charge degeneracy through the coupling to the auxiliary QD. Since the normal and anomalous tunneling processes between the chain and the additional QD are proportional to $u$ and $v$, respectively, the resulting anticrossing amplitudes become parity dependent. As a consequence, the maximum QC values differ in the even and odd sectors, as described by Eqs.~(\ref{Ce}) and~(\ref{Co}).

Next, we consider a situation in which the even and odd ground-state energies differ while the MBSs remain spatially separated ($|M| = 1$). The energy splitting shifts the position of the anticrossings in each parity sector, as illustrated in Fig.~\ref{Fig2}(e) for $|\delta E_0| = 0.2$. This shift occurs because the avoided crossing occurs whenever the auxiliary QD and the Kitaev chain become resonant in each sector. This is also explicit in Eqs.~(\ref{Ce}) and~(\ref{Co}), where the $\varepsilon_d$ dependency of $C_{e(o)}$ is shifted by the hybridization between the MBSs. The resulting shift in the anticrossings is reflected in the QC as split peaks, as shown in Fig.~\ref{Fig2}(f), measurable through $\delta E_0 = \delta/2$. 

In the most general situation, both energy splitting and MBS overlap are present. To more systematically analyze this case, we plot $\nu$ and $\delta$ over the full parameter space of the Kitaev chain by varying $\varepsilon_{2,3}$ and $t_{2,3}$. Figures~\ref{Fig2}(g,h) show that energy splitting and MBS overlap can be independently measured from QC, extending the findings shown in Fig.~\ref{Fig1}(d) to a broader context. The results in Fig.~\ref{Fig2}(g,h)  display white lines that identify optimal values of the energy splitting ($\delta E_0 = 0$) and MP ($|M| = 1$), see the dashed lines in Figs.~\ref{Fig2}(g,h). The MBS sweet spot appears as the intersection of these lines~\cite{tsintzis2022creating, Tsintzis2024, Dourado_Kitaev3}, see Sec.~A of the supplemental material (SM)~\cite{SM} for a more detailed comparison between analytic and numerical results.

In Sec.~B of the SM, we show that qualitatively similar results are obtained interactions are considered in the QDs, demonstrating that MBS sweet spots appearing at sufficiently large magnetic fields can be identified with QC. In Sec.~C of the SM, we show that the same conclusions hold if we consider charging energy between the system and the auxiliary QD. In Sec.~D of the SM, we relax the symmetry requirement and show that an auxiliary QD on the right side of the Kitaev chain can be used to assess the right MP, $|M_3|$, while still relying exclusively on QC measurements from the left side.

\begin{figure}
\centering
\includegraphics[width=\linewidth]{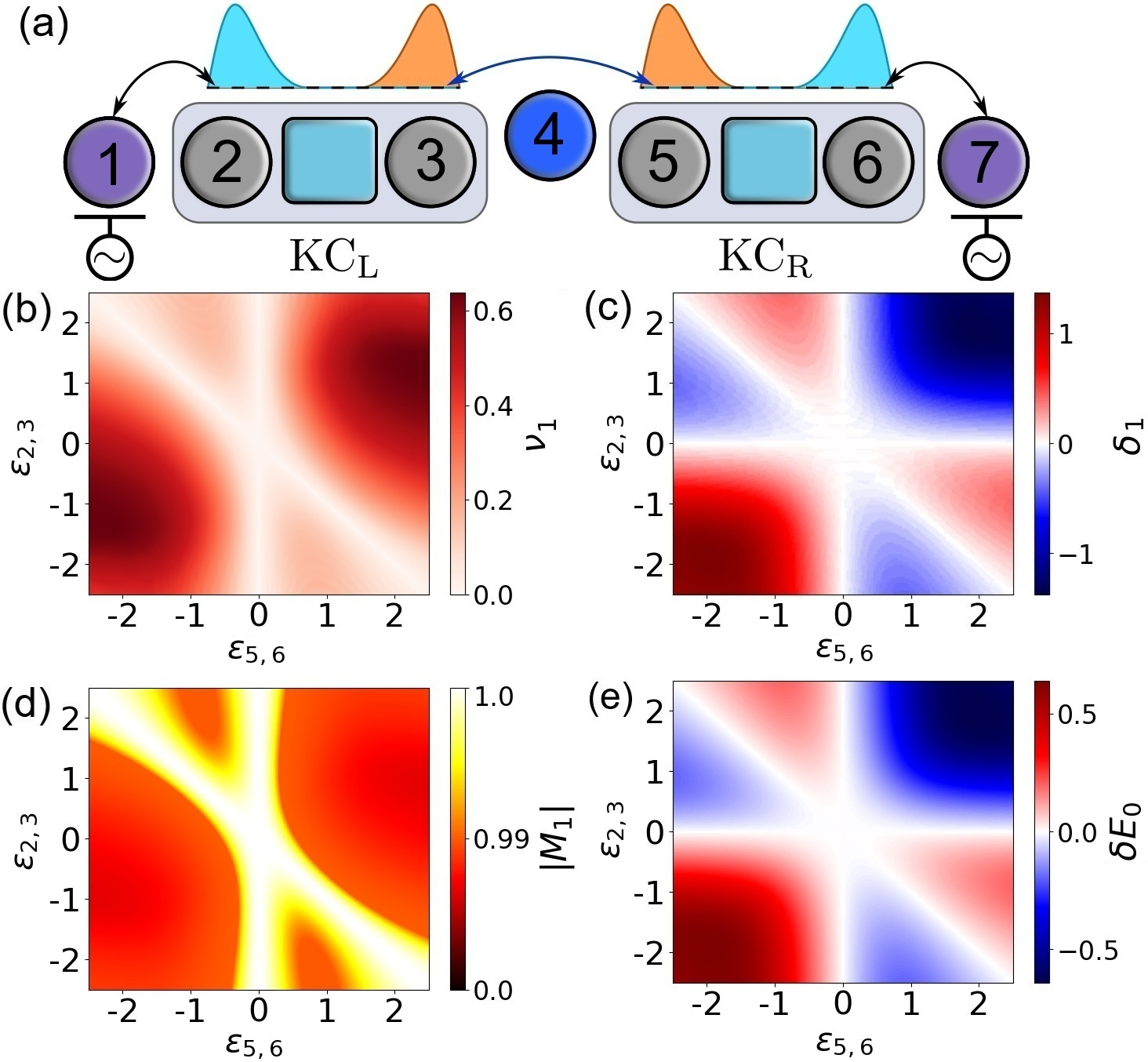}
\caption{Tuning a MBS-based qubit via QC. (a) Sketch of a MBS qubit in two minimal Kitaev chains, $KC_L$ and $KC_R$, are coupled through a central QD (blue) and to auxiliary QDs (purple). QC measurements at the outer QDs enable tuning to the operational sweet spot. (b),(c) $\nu_1$ and $\delta_1$ as functions of $\varepsilon_{2,3}$ and $\varepsilon_{5,6}$, respectively. (d),(e) $|M_1|$ and $\delta E_0$ as functions of of $\varepsilon_{2,3}$ and $\varepsilon_{5,6}$, respectively. Parameters: $t_{1,2} = t_{6,7} = 0.5$, $t_{2,3} = t_{3,4} = t_{4,5} = t_{5,6} = 1$, and $\varepsilon_4 = 0$.}
\label{Fig3}
\end{figure}

\textit{Assessing the quality of MBS qubits.} We now turn to a qubit setup, where two systems hosting isolated MBSs couple. As an example, we will focus on a qubit based on minimal Kitaev chains~\cite{Pino_PRB2024,Pan_PRB2025}, shown in Fig.~\ref{Fig3}(a). The device includes a central QD to control and drive the coupling between the Kitaev chains, and two auxiliary QDs on the left and right sides, which are in turn coupled to resonators to measure QC. The corresponding Hamiltonian for the Kitaev chain-based qubit is given by
\begin{equation}
H_{Q} = \sum_{i=1}^7 \varepsilon_i c_i^\dagger c_i + \sum_{i=1}^6 \left[ t_{i, i+1} c_{i}^\dagger c_{i+1} + \Delta_{i, i+1} c_{i}^\dagger c_{i+1}^\dagger + \text{H.c.} \right],
\end{equation}
where $\Delta_{i,i+1} = 1$ only for $i=2,5$. 

When the central QD, shown in blue in Fig.~\ref{Fig3}(a), is tuned near resonance, the inner MBSs [orange blobs in Fig.~\ref{Fig3}(a)], hybridize, lifting the initial four-fold ground state degeneracy, leading to two sets of levels, one per total parity. 
In this regime, the QC measured from the auxiliary QDs can be used to estimate the local MP at the ends of the system ($|M_2|$ and $|M_6|$), as well as the energy splitting between the total even- and odd-parity states. To corroborate the discussion above, we investigate the QC signals as the site energies are varied in the two Kitaev chains. In Figs.~\ref{Fig3}(b,c), we plot $\nu_1$ and $\delta_1$ (both measured at the auxiliary QD $1$) as functions of $\varepsilon_{2,3}$ and $\varepsilon_{5,6}$, respectively. We note that the white lines, corresponding to $\nu_1 = 0$ and $\delta_1 = 0$, indeed coincide with $|M_1| = 1$ and $\delta E_0 = 0$, respectively, as shown in Figs.~\ref{Fig3}(d,e). By imposing $\nu_{1,7} = 1$ and $\delta = 0$, the conditions for well-localized MBSs at the ends of the qubit and degenerate ground states are satisfied, which are necessary conditions for a protected MBS-based qubit. 

We note, however, two configurations where the system features high MP at the left side and degenerate ground states: $\varepsilon_{5,6}=0$ ($\varepsilon_{2,3} \neq 0$) and $\varepsilon_{5,6} = -\varepsilon_{2,3}$, white lines in Figs.~\ref{Fig3}(b). The first condition corresponds to delocalizing the outermost MBSs [blue in Fig.~\ref{Fig3}(a)] toward the middle of the system. This situation leads to a reduction in the local MP on the right side ($|M_6| < 1$) that can be detected by measuring the QC on the right side ($\nu_7$). On the other hand, the line at $\varepsilon_{2,3} = -\varepsilon_{5,6}$ is an accidental degeneracy and can be lifted by varying the energy of the central QD, $\varepsilon_4$. This is confirmed in Sec.~E of the SM, where we also demonstrate that the MP and energy splitting remain accessible through QC measurements across different parameter regimes, including weaker inter-chain couplings.

Therefore, it is possible to extract the individual MPs and energy splittings from QC measured at the two ends of a MBS-based qubit. However, assessing the MBS quality at the middle is more challenging and would require performing QC measurements at the central QD in a configuration where the QD only couples to one half of the system.

\textit{Conclusions}. In this work, we have introduced a scheme to characterize MBSs using QC measurements. We have shown that the QC signals in the even and odd parity sectors differ when MBSs overlap or when there is a finite energy splitting between the ground states of the two parity sectors. We demonstrated that QC measurements are sufficient to estimate both the local MBS overlap and the energy splitting, enabling, e.g., tuning QD-based Kitaev chains to MBS sweet spots. We also considered a qubit geometry in which two Kitaev chains are coupled through a QD. Deviations from the ideal MBS condition in either or both chains can be detected via the QC of the QDs at the ends of the system.

Experimentally, the QC maxima in the two parity sectors can be obtained with either one or two measurements, depending on whether the quasiparticle poisoning time is shorter or longer than the QD sweep time. Characterizing Kitaev chains and MBS-based qubits without relying on electron spectroscopy techniques is crucial for future experiments aiming to demonstrate the non-Abelian properties of MBSs, where quasiparticle poisoning is detrimental.

During the final stages of preparation of the present manuscript, we became aware of Ref.~\cite{aghaee202620secondparitylifetime}, which reports QC measurements using dispersive gate sensing through auxiliary QDs, similar to the ideas discussed here.

\textit{Acknowledgments}. R.A.D. acknowledges financial support from Fundação de Amparo à Pesquisa do Estado de São Paulo (FAPESP), Brazil (Grant No. 2025/18261-2). M.L. acknowledges funding from the European Research Council (ERC) under the European Unions Horizon 2020 research and innovation programme under Grant Agreement No. 856526, the Swedish Research Council under Grant Agreement No. 2024-05491, and NanoLund. R.A. acknowledges funding from the Horizon Europe Framework Program of the European Commission through the European Innovation Council Pathfinder Grant No. 101115315 (QuKiT). R.S.S acknowledges funding from the Spanish Comunidad de Madrid (CM) ``Talento Program'' (Project No. 2022-T1/IND-24070).
R.A. and R.S.S acknowledge the Spanish Ministry of Science, Innovation, and Universities through Grants CEX2024-001445-S (Severo Ochoa Centres of Excellence program), PID2022-140552NA-I00 and PID2024-161156NB-I00.  
\bibliography{bibliography.bib}

\end{document}


\preprint{APS/123-QED}

\title{Supplemental Material for ``Assessing Majorana states and qubits through quantum capacitance"}

\author{Rodrigo A. Dourado}
\email{dourado.rodrigo.a@gmail.com}
\affiliation{ \textit{Instituto de F\'{ı}sica de S\~{a}o Carlos, Universidade de S\~{a}o Paulo, 13560-970 S\~{a}o Carlos, S\~{a}o Paulo, Brazil}	}

\author{Ramón Aguado}
\affiliation{Quantum Advanced Research Center (QuARC), Consejo Superior de Investigaciones Científicas (CSIC), Sor Juana Inés de la Cruz 3, 28049 Madrid, Spain}
\affiliation{Instituto de Ciencia de Materiales de Madrid (ICMM), Consejo Superior de Investigaciones Cient\'{i}ficas (CSIC), Sor Juana In\'{e}s de la Cruz 3, 28049 Madrid, Spain}

\author{Jeroen Danon}
\affiliation{Department of Physics, Norwegian University of Science and Technology, Trondheim NO-7491, Norway}

\author{Martin Leijnse}
\affiliation{Division of Solid State Physics and NanoLund, Lund University, S-22100 Lund, Sweden}

\author{Rub\'{e}n Seoane Souto}
\affiliation{Quantum Advanced Research Center (QuARC), Consejo Superior de Investigaciones Científicas (CSIC), Sor Juana Inés de la Cruz 3, 28049 Madrid, Spain}
\affiliation{Instituto de Ciencia de Materiales de Madrid (ICMM), Consejo Superior de Investigaciones Cient\'{i}ficas (CSIC), Sor Juana In\'{e}s de la Cruz 3, 28049 Madrid, Spain}

\date{\today}

\maketitle

\section{Correspondence between measurement and theory} \label{CorrespondenceApp}

\begin{figure}
\centering
\includegraphics[width=0.6\linewidth]{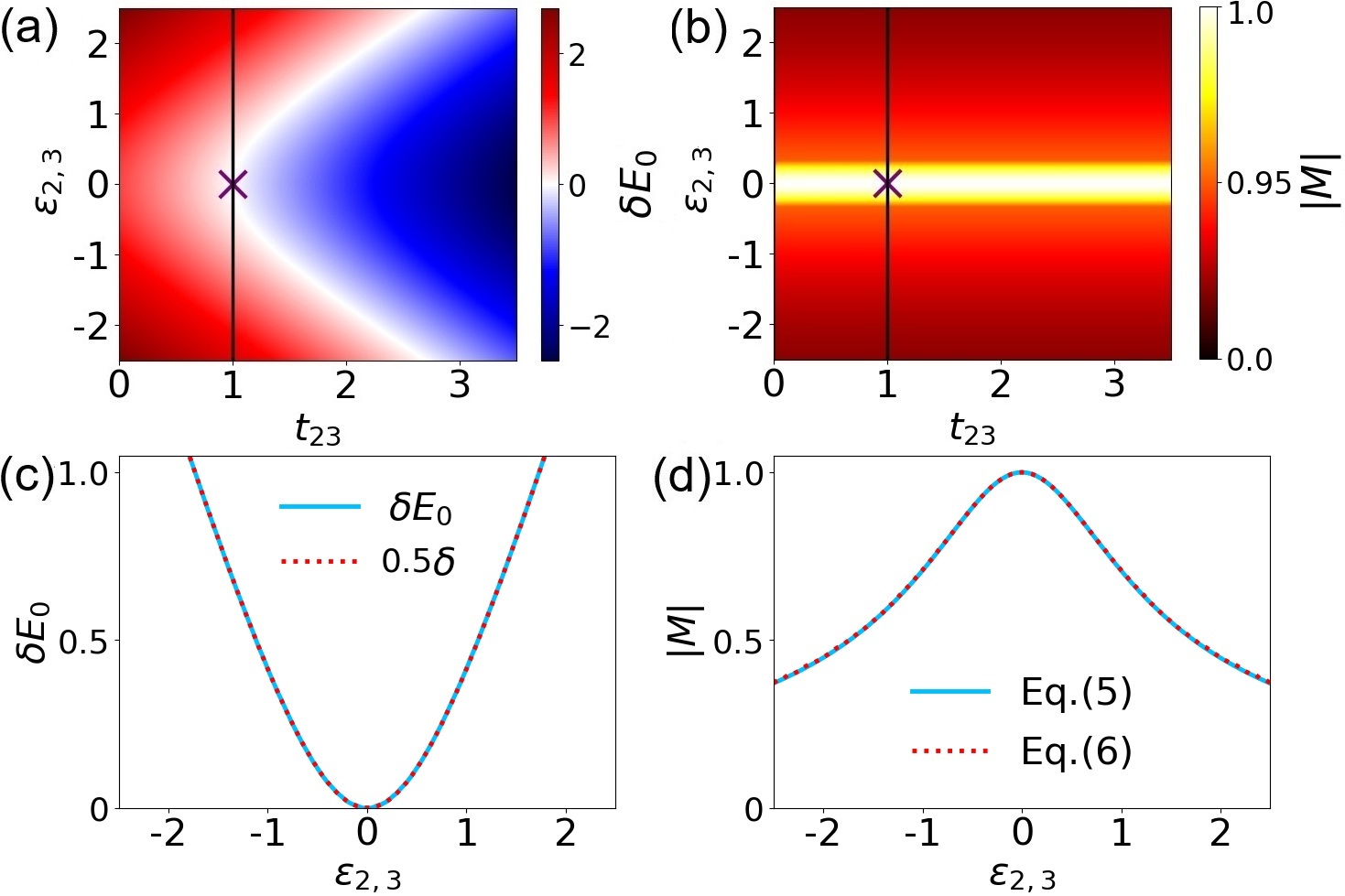}
\caption{Identifying MBS sweet spots in parameter space via QC.
(a),(b) Energy splitting $\delta E_0$ and MP $|M|$ as functions of $\varepsilon_{2, 3}$ and $t_{2,3}$. (c),(d) Quantitative agreement between $\delta E_0$ (solid blue) and $2\delta$ (dotted red), and between the MP calculated from Eq.~(5) (solid blue) and measured from Eq.~(6) (dotted red).}
\label{FigA1}
\end{figure}

In this section, we compare the expressions to estimate the ground state energy splitting and Majorana polarization (MP) through quantum capacitance (QC) measurements to the values calculated from the definitions. To this end, we start by plotting $\delta E_0$ and $|M|$ over the full parameter space of the Kitaev chain and compare to the results shown in Figs.~2(g,h). The results in Figs.~\ref{FigA1}(a,b) display white lines that identify optimal values of the energy splitting and MP. The MBS sweet spot appears as the intersection of these lines (marked by crosses), where $\delta E_0 = 0$ and $|M| = 1$~\cite{tsintzis2022creating, Tsintzis2024, Dourado_Kitaev3}. Qualitatively, the same features emerge as in Figs.~2(g,h): white lines corresponding to $\delta = 0$ and $\nu = 0$. Their intersection, $\delta = \nu = 0$, coincides with the crosses in Figs.~\ref{FigA1}(a,b), located at $\varepsilon_{2, 3} = 0$ and $t_{2,3} = 1$. 

To verify the quantitative agreement between the measured quantities and the exact values calculated from the definitions, we vary $\varepsilon_{2,3}$ while fixing $t_{2,3} = 1$ [black lines in Figs.~\ref{FigA1}(a,d)] and plot the corresponding quantities in Figs.~\ref{FigA1}(c,d). We find excellent agreement between $\delta/2$ and $\delta E_0$ in Fig.~\ref{FigA1}(c), as well as between the MP in the definition of Eq.~(5) and the one obtained from $\nu$, Eq.~(6). We conclude that $\delta E_0$ and $|M|$ can be directly estimated from QC measurements, which suffices to tune the system to the two-site Kitaev sweet spot.

\section{Interacting model} \label{interactingModel}

In this section, we corroborate the findings discussed in the main text using a model of interacting quantum dots (QDs), see Fig.~1(e), at a finite magnetic field. The system consists of normal QDs coupled via narrow superconductors that host subgap states, modeled as QDs with induced superconducting correlations. We consider an external magnetic field that polarizes the spins in the QDs. The Hamiltonian that describes the QDs is given by

\begin{equation}
    H_{\rm QDs} = \sum_{i, \sigma}^4 (\varepsilon_{i} + s_\sigma \varepsilon_{z, i}) n_{i, \sigma} +\sum_i^4 U_i n_{i, \uparrow}n_{i, \downarrow}+
    \left(\Delta d_{3, \uparrow}^\dagger d_{3, \downarrow}^\dagger + {\rm H.c.} \right),
    \label{Eq:Hmicroscopic}
\end{equation}
which describes an effective Kitaev chain with $2$ sites coupled on the left side to an auxiliary QD. Here, $\varepsilon_{i}$ is the on-site energy of QD $i$, $\varepsilon_z$ is the Zeeman energy ($s_{\uparrow/\downarrow} = \pm 1$), and  $U_i$ is the local electrostatic repulsion, with $n_{i,\sigma}=d^\dagger_{i,\sigma} d_{i,\sigma}$ being the number operator and $d_{i,\sigma}$ the annihilation operator for electrons with spin $\sigma=\uparrow,\downarrow$ at site $i$. We describe the coupling between QDs and the superconductor as an induced pairing amplitude $\Delta$.

The tunneling between the QDs is given by
\begin{equation} \label{TunnelingQDs}
    H_T= \sum_{i, \sigma}^{3}\left[ t_i d_{i+1, \sigma}^\dagger d_{i, \sigma} + t_{i}^{so} s_{\sigma}d_{i+1, \sigma}^\dagger d_{i, \bar{\sigma}}+ {\rm H.c.} \right],
\end{equation}
where $t$ and $t^{\mathrm{so}}$ are the spin-conserving and spin-flip tunneling amplitudes, respectively, with $t^{\mathrm{so}}$ arising from spin-orbit coupling and $\bar{\sigma}$ denoting the spin opposite to $\sigma$. The total Hamiltonian is $H = H_{\rm QDs} + H_T$. We consider that screening effects in the superconductor suppress the applied magnetic field and the Coulomb interaction, such that $\varepsilon_{z, 3} = U_3 = 0$. Besides that, we consider $\varepsilon_{z, i} = 1.5\Delta$ and $U_i = 5\Delta$ for $i = 1,2,4$, $t_1 = 0.01\Delta$, $t_i = 0.5\Delta$, $i=2, 3, 4$, and $t_i^{so} = 0.2 t_i$.

\begin{figure}
\centering
\includegraphics[width=0.6\linewidth]{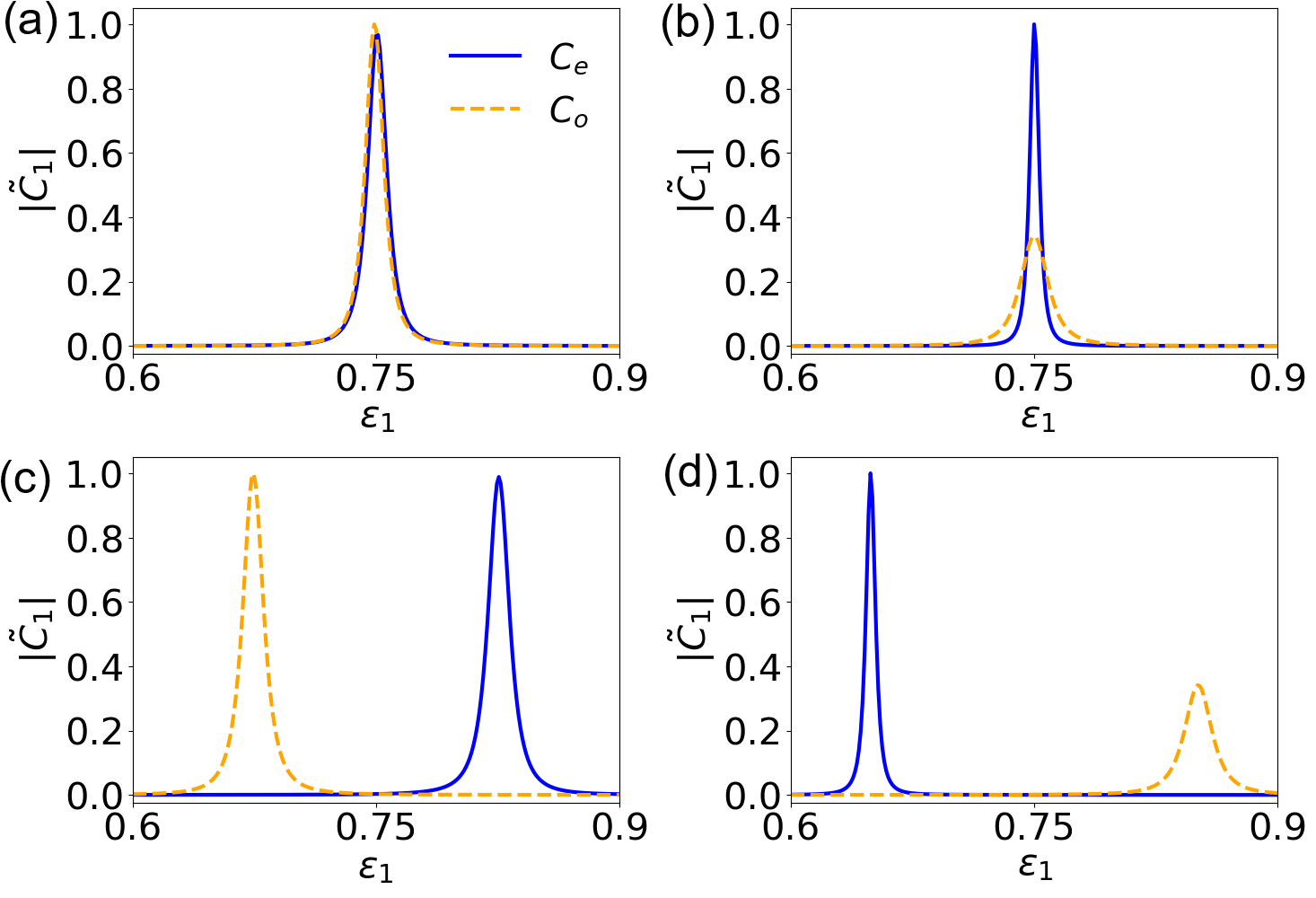}
\caption{Interacting model. QC as a function of $\varepsilon_1$ for (a) A sweet spot, $|M_2| = 0.985$, $|\delta E_0| = 0$, (b) Delocalized MBSs, $|M_2| = 0.49$, $|\delta E_0| = 0$, (c) Energy splitted ground states, $|M_2| = 0.99$, $|\delta E_0| = 0.075$, and (d) Delocalized MBSs with splitted ground states, $|M_2| = 0.59$, $|\delta E_0| = 0.1$.} 
\label{FigC1}
\end{figure}

The above-described Hamiltonian is capable of hosting well-isolated MBSs in the appropriate regime, although for finite magnetic fields, there will be a residual overlap between the MBSs at the edges, resulting in $|M_{2, 4}| < 1$. For the considered parameters, the best achievable MP while also ensuring degeneracy of the ground states is $~0.985$~\cite{tsintzis2022creating}. We tune the QDs to this sweet spot and vary the auxiliary QD level while calculating the QC in Fig.~\ref{FigC1}(a). As discussed in the main text, the curves for even and odd ground states are nearly identical, with a small difference in the peak height due to $|M_2| < 1$. We note that the peaks are not centered around $\varepsilon_1 = 0$, but rather $\varepsilon_1 = \varepsilon_z/2$, and another resonance occurs at $\varepsilon_1 = -\varepsilon_z/2 - U$. These two values make the auxiliary QD resonant with the Fermi level. 

Next, we consider deviations from the sweet spot. For delocalized MBSs, see Fig.~\ref{FigC1}(b), we note a more pronounced peak height difference between even and odd sectors, similar to Fig.~2(d). Also similar to the Kitaev model, an energy splitting is reflected in split QC curves for even and odd states, for localized and delocalized MBSs, as shown in Figs~\ref{FigC1}(c,d). These results confirm that the findings explored in the main text are valid in the presence of interactions and finite magnetic fields. 

\section{Charging energy between the system and the auxiliary QD}

\begin{figure}
\centering
\includegraphics[width=0.95\linewidth]{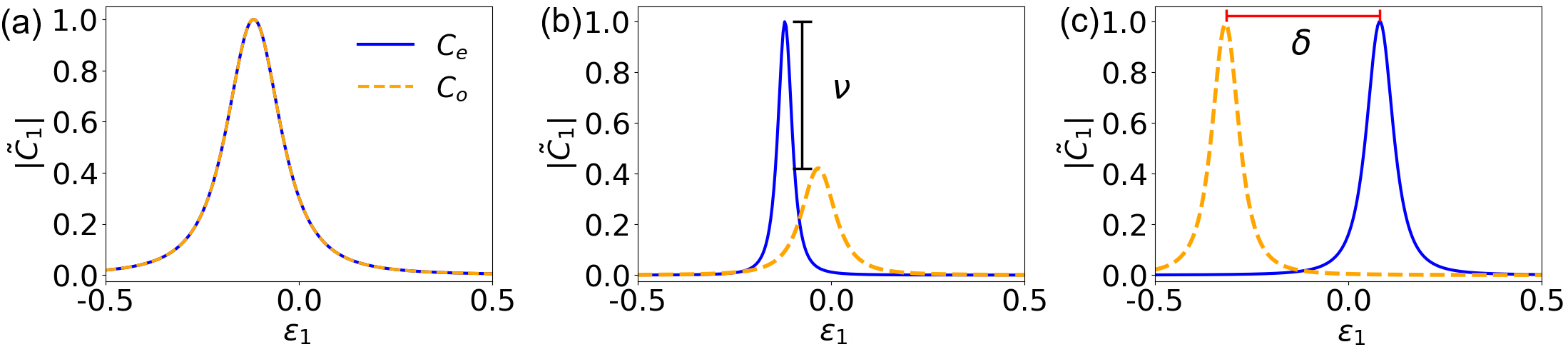}
\caption{QCs as a function of the auxiliary QD level $\varepsilon_1$ for (a) the MBS sweet spot ($|\delta E_0| = 0$, $|M| = 1$), (b) poorly localized MBSs ($|\delta E_0| = 0$, $|M| = 0.7$), and (c) for finite energy splitting ($|\delta E_0| = 0.2$, $|M| = 1$). The parameters are the same as the ones in Fig.~2, with $U_d = 0.25$.} 
\label{FigUd}
\end{figure}

In this section, we consider charging energy between the auxiliary QD and the system. Similar to projecting the local fermionic operator coupled to the auxiliary QD onto the low-energy fermion $f$ as approximately
$u f + v f^\dagger$, the Coulomb interaction between the auxiliary QD dot and the first site of the system becomes $U_d|v|^2 n_d + U_d\left(|u|^2-|v|^2\right)n_d n_f$, where $n_\alpha = \alpha^\dagger \alpha$. The effective Hamiltonian then reads

\begin{equation}
H_{\rm eff}
=
\varepsilon_d n_d+\varepsilon_M n_f
+t d^\dagger(uf+vf^\dagger)+\mathrm{H.c.}
+U_d|v|^2 n_d
+U_d\left(|u|^2-|v|^2\right)n_d n_f.
\end{equation}

In the even basis $\{|0\rangle,d^\dagger f^\dagger|0\rangle\}$ and odd basis
$\{f^\dagger|0\rangle,d^\dagger|0\rangle\}$, the Hamiltonians become

\begin{equation}
H_e=
\begin{pmatrix}
0 & tv \\
t^*v^* & \varepsilon_d+\varepsilon_M+U_d|u|^2
\end{pmatrix},
\qquad
H_o=
\begin{pmatrix}
\varepsilon_M & tu \\
t^*u^* & \varepsilon_d+U_d|v|^2
\end{pmatrix}.
\end{equation}

The corresponding ground state energies are

\begin{equation}
E_0^e=
\frac{\varepsilon_d+\varepsilon_M+U_d|u|^2}{2}
-\frac12
\sqrt{
(\varepsilon_d+\varepsilon_M+U_d|u|^2)^2+4|tv|^2
},
\end{equation}

\begin{equation}
E_0^o=
\frac{\varepsilon_d+\varepsilon_M+U_d|v|^2}{2}
-\frac12
\sqrt{
(\varepsilon_d-\varepsilon_M+U_d|v|^2)^2+4|tu|^2
},
\end{equation}
and the corresponding QCs, $C_q^{e,o}=-\partial^2 E_0^{e,o}/\partial\varepsilon_d^2$, are

\begin{equation}
C_q^e=
\frac{2|tv|^2}
{\left[(\varepsilon_d+\varepsilon_M+U_d|u|^2)^2+4|tv|^2\right]^{3/2}},
\qquad
C_q^o=
\frac{2|tu|^2}
{\left[(\varepsilon_d-\varepsilon_M+U_d|v|^2)^2+4|tu|^2\right]^{3/2}}.
\end{equation}

The QCs maxima occur at

\begin{equation}
    \varepsilon_d^{(e)}=-\varepsilon_M-U_d|u|^2,
    \qquad
    \varepsilon_d^{(o)}=\varepsilon_M-U_d|v|^2.
\end{equation}
We note that $U_d$ does not affect the peak heights for the QCs, such that the measurement of the MP through $\nu$, see Eq.~(6) in the main text. In fact, the charging energy between the auxiliary QD and the system makes the QC curves more sensitive to deviations from $u = v$, as a overlapping MBSs lead to an additional shift on the QC maxima, which is given by $\delta = 2 \varepsilon_M + U_d \left(|u|^2 - |v|^2 \right)$. Therefore, by minimizing both $\delta$ and $\nu$, one can tune to well-localized zero-energy MBSs even in the presence of charging energy between the auxiliary QD and the system. 

Next, we reproduce the QC curves shown in Figs.~1(b,d,f) using the same parameters and including interdot charging energy, given by the term $U_d c^\dagger_1 c_1 c^\dagger_2 c_2$ to the Hamiltonian described by Eq.~(7) in the main text. In Fig.~\ref{FigUd}(a), we observe that in agreement with the above discussion, $C_e = C_o$ for the MBS sweet spot, where the energy splitting vanishes and $|M| = 1$ ($|u| = |v|$). For poorly localized MBSs, the value of $\nu$ remains the same as in Fig.~1(d), but a split $\delta$ arises as $|u| \neq |v|$, see Fig.~\ref{FigUd}(b). For localized MBSs ($|M| = 1$) with energy-splitted ground states as in Fig.~1(f), the measurement is not affected by the charging energy, as both QC curves are equally shifted by $U_d$, see Fig.~\ref{FigUd}(c).

\section{Asymmetric Kitaev chains} \label{asymmetric chains}

As in experiments, it is not possible to ensure the symmetry $|M_2| = |M_3|$. We therefore consider the case $\varepsilon_2 \neq 0$, such that $|M_3| < 1$. It has been predicted~\cite{Seoane2023} and experimentally verified~\cite{Bordin_arXiv2025} that coupling a QD to overlapping MBSs leads to energy splittings in the spectrum. To probe the MBS overlap at site 3, we introduce a resonant additional QD coupled to the right side of the Kitaev chain, see Fig.~\ref{FigC1}(a). In this configuration, when $|M_3| < 1$, the resulting energy splitting can be detected via the leftmost QD, as shown in Fig.~2(e).

\begin{figure}
\centering
\includegraphics[width=0.6\linewidth]{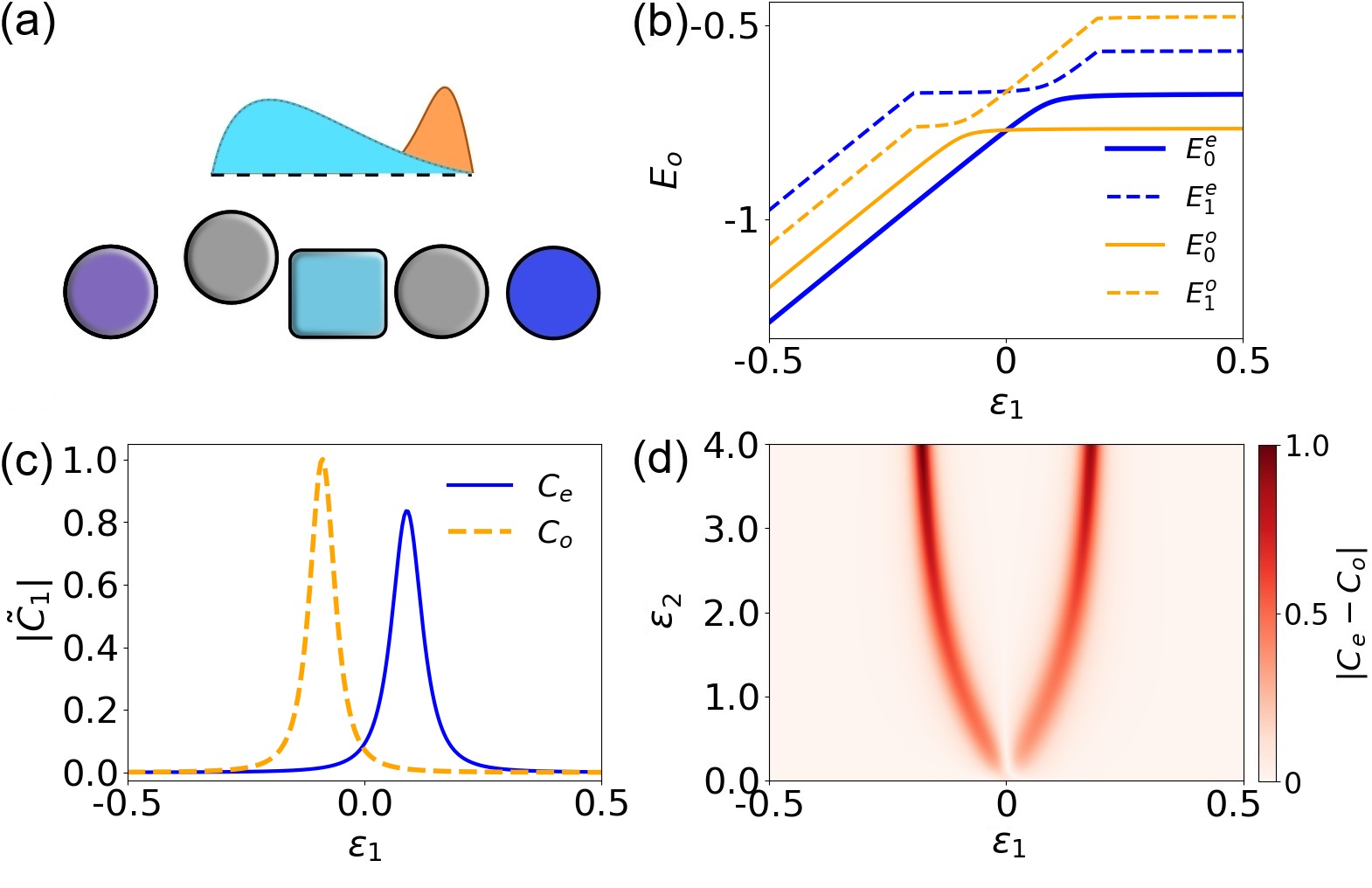}
\caption{Detecting MP in asymmetric Kitaev chains.
(a) Asymmetric Kitaev chain, where detuning a QD induces MBS overlap on the right side. (b) Energy spectrum in the even and odd parity sectors as a function of the left QD level $\varepsilon_1$. (c) QC as a function of $\varepsilon_1$ for even and odd parity sectors, with $|M_2| = 1$ and $|M_3| = 0.7$. (d) $|C_{1,e} - C_{1,o}|$ as a function of $\varepsilon_2$ and $\varepsilon_1$. Parameters: $\varepsilon_4 = 0$, $t_{12} = 0.05$, $t_{23} = 1$, and $t_{34} = 0.1$.} 
\label{FigB1}
\end{figure}

To corroborate this picture, we set $|M_3| = 0.7$ ($|M_2| = 1$) and plot the energies and QC as functions of $\varepsilon_1$. As expected, an energy splitting emerges due to the MBS overlap at site 3, as shown in Fig.~\ref{FigB1}(b), leading to different positions of the anticrossings in the even and odd sectors. This, in turn, results in split QC peaks, see Fig.~\ref{FigB1}(c), similar to Fig.~2(e), with the distinction that the splitting now originates from the coupling between the two MBSs to the rightmost QD. For a more systematic analysis, Fig.~\ref{FigB1}(d) shows that as $\varepsilon_2$ increases—thereby reducing $|M_3|$ while keeping $|M_2| = 1$, the splitting in the QC increases and eventually saturates around $\varepsilon_2 \approx 4$, where $|M_3|$ becomes vanishingly small. 

\section{Additional plots for the qubit setup}

\begin{figure}
\centering
\includegraphics[width=0.6\linewidth]{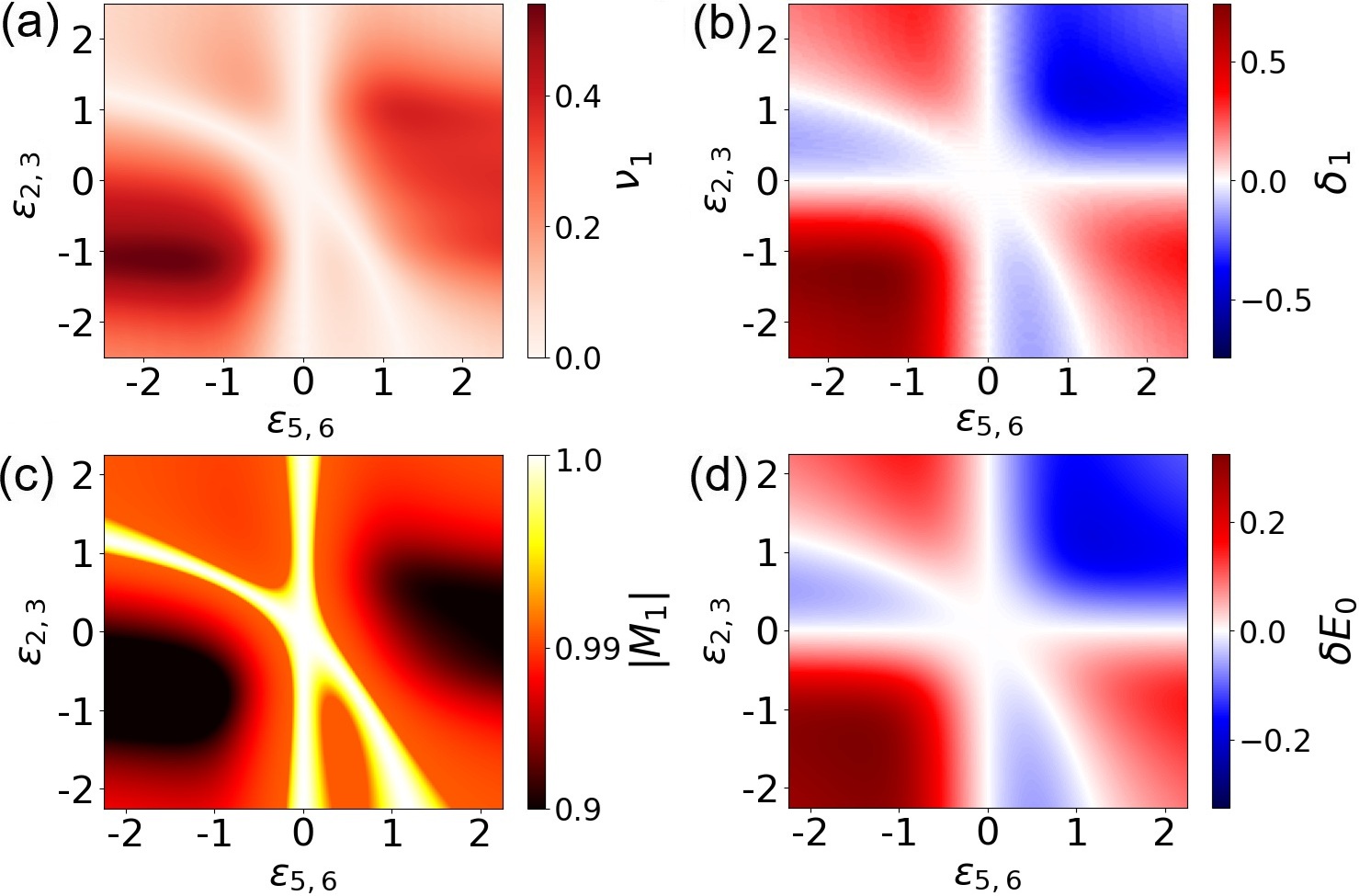}
\caption{(a),(b) $\nu_1$ and $\delta_1$ as functions of $\varepsilon_{2,3}$ and $\varepsilon_{5,6}$, respectively. (c),(d) $|M_1|$ and $\delta E_0$ as functions of of $\varepsilon_{2,3}$ and $\varepsilon_{5,6}$, respectively. The p arameters are the same as in Fig.~3, apart from $t_{3,4} = t_{4,5} = 0.5$, and $\varepsilon_4 = 0.1$.}
\label{FigD1}
\end{figure}

\begin{figure}
\centering
\includegraphics[width=0.6\linewidth]{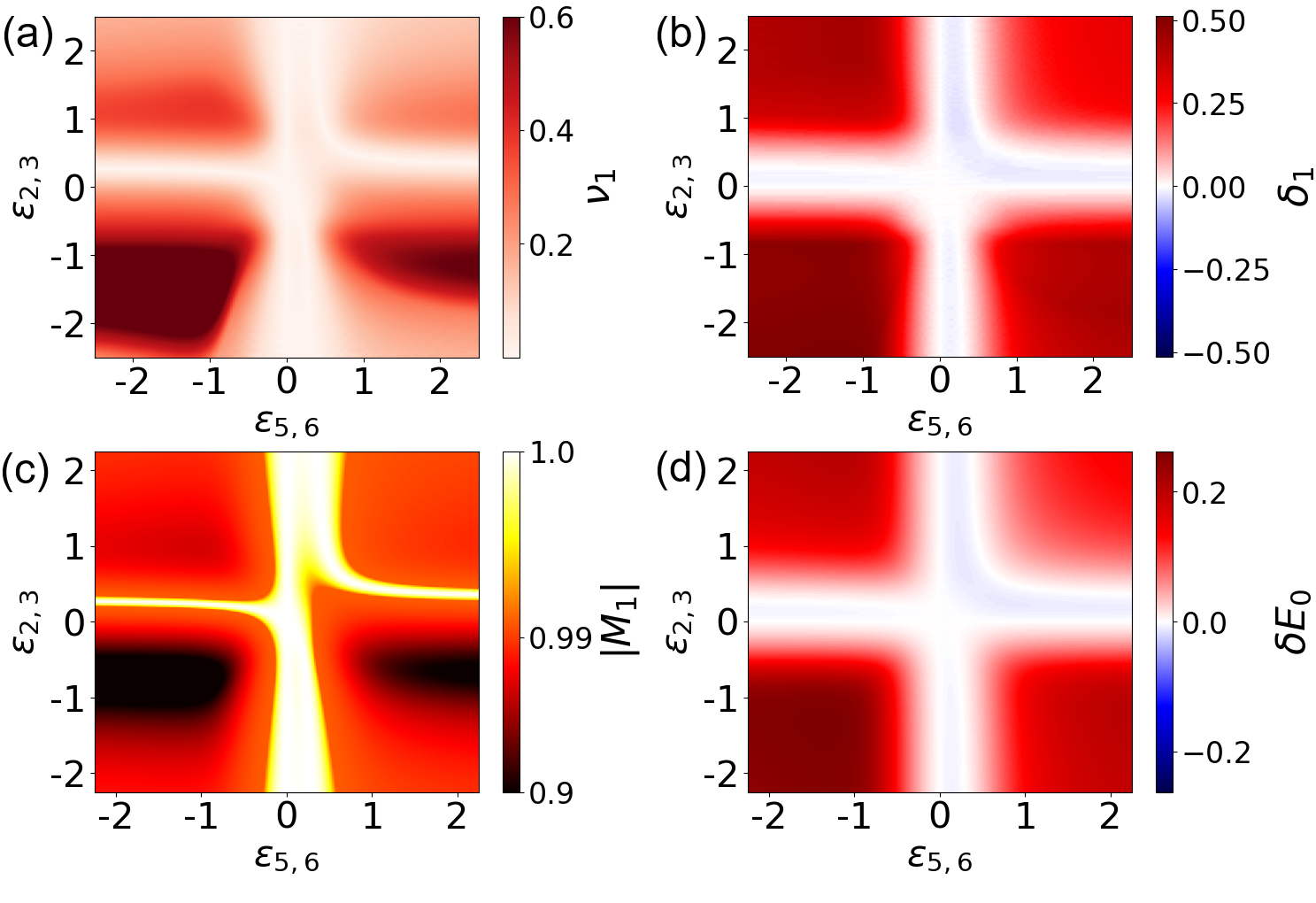}
\caption{(a),(b) $\nu_1$ and $\delta_1$ as functions of $\varepsilon_{2,3}$ and $\varepsilon_{5,6}$, respectively. (c),(d) $|M_1|$ and $\delta E_0$ as functions of of $\varepsilon_{2,3}$ and $\varepsilon_{5,6}$, respectively. The p arameters are the same as in Fig.~3, apart from $t_{3,4} = t_{4,5} = 0.25$, and $\varepsilon_4 = 0.2$.}
\label{FigD2}
\end{figure}

In this section, we explore different parameters for the qubit setup. We focus on detuning the central QD and reducing its coupling to the Kitaev chains. In Fig.~\ref{FigD1}, we consider the case where $\varepsilon_d = 1$ and $t_{3,4} = t_{4,5} = 0.5$, and in Fig.~\ref{FigD2}, $\varepsilon_d = 0.2$ and $t_{3,4} = t_{4,5} = 0.25$. We note that the line corresponding to high $|M_1|$ and $\delta E_0 = 0$ for $\varepsilon_{2, 3} = -\varepsilon_{5,6}$ bends as the coupling between the two Kitaev chains is reduced. However, the correspondence between $\nu_1$ and $|M_1|$ and $\delta$ and $\delta E_0$ remains true as we vary the coupling between the Kitaev chains.  

\bibliography{bibliography.bib}